\documentclass[10pt,twocolumn]{article}

\usepackage{multicol}
\usepackage{graphicx}
\usepackage{amsmath}
\usepackage{amssymb}
\usepackage{lipsum}
\usepackage{color}
\usepackage{siunitx}
\usepackage{authblk}
\usepackage[margin=2cm]{geometry}
\usepackage{subeqnarray}

\title{Diffusiophoretic manipulation of particles in a drop deposited on a hydrogel}
\author[1,2,3]{Fran\c{c}ois Boulogne}
\author[2,4]{Sangwoo Shin}
\author[1]{Julien Dervaux}
\author[1]{Laurent Limat}
\author[2]{Howard A. Stone}
\affil[1]{{\small Laboratoire Mati\`ere et Syst\`emes Complexes (MSC), UMR 7057 CNRS, Universit\'e Paris Diderot, B\^atiment Condorcet, 10 rue Alice Domon et L\'eonie Duquet, Paris, France}}
\affil[2]{{\small Department of Mechanical and Aerospace Engineering, Princeton University, Princeton, NJ 08544, USA}}
\affil[3]{{\small Laboratoire de Physique des Solides, CNRS, Univ. Paris-Sud, Universit\'e Paris-Saclay, Orsay 91405, France}}
\affil[4]{{\small Department of Mechanical Engineering, University of Hawaii at Manoa, Honolulu, HI 96822, USA}}
\date{\today}
\begin{document}


\twocolumn[
    \begin{@twocolumnfalse}
        \maketitle
        \begin{abstract}
            We report an experimental study on the manipulation of colloidal particles in a drop sitting on a hydrogel.
            The manipulation is achieved by diffusiophoresis, which describes a directed motion of particles induced by solute gradients. By letting the solute concentrations for the drop and the hydrogel be different, we control the motion of particles in a stable suspension, which is otherwise difficult to achieve.
            We show that diffusiophoresis can cause the particles to move either toward or away from the liquid-air interface depending on the direction of the solute gradient and the surface charge of the particles.
            We measure the particle adsorption experimentally and rationalize the results with a one-dimensional numerical model.
            We show that diffusiophoretic motion is significant at the lengthscale of a drop deposited on a hydrogel, which suggests a simple method for the deposition of particles on hydrogels.
        \end{abstract}
    \end{@twocolumnfalse}
]

\section{Introduction}

The transport of colloidal particles is of crucial interest for microfabrication, self-assembly and surface coating \cite{Whitesides2002,Whitesides2002a,Lin2003,Grzelczak2010,Merlin2012}.
A broad range of methods are available to manipulate nano- to micrometer size particles.
One of the most common mechanisms for colloidal transport relies on solvent evaporation.
Therefore, drying receives significant attention from the evaporation of dilute colloidal droplets to the consolidation of thin films of concentrated solutions (see \cite{Larson2013,Routh2013} for reviews).
Nevertheless, evaporation is particularly difficult to control both in time and space.
To mitigate coating inhomogeneities, different strategies can be employed such as the Marangoni effect \cite{Poulard2007,Kajiya2009,Still2012,Kim2016}, flow properties modified by solvent volatility \cite{Kim2016} or convection \cite{Selva2012}.

Ultimately, to have a good control over the coating process, manipulating the motion of particles via external forces is required.
Indeed, the idea to apply forces on particles is long standing.
Phoretic phenomena have been known for decades to induce particle motion \cite{Derjaguin1961,Dukhin1974,Anderson1984,Anderson1989} and they recently regained interest either for particle manipulation \cite{Abecassis2008,Abecassis2009,Kar2015,Shin2016,Shi2016} or to explain an exclusion zone adjacent to solids \cite{Florea2014} (see \cite{Velegol2016} for a recent review).
In particular, Derjaguin, Dukhin and Korotkova revealed the role of diffusiophoresis in the deposition of latex particles for rubber film formation \cite{Derjaguin1961}. 
This method has been used since the 1920's \cite{Prieve1982}.
More specifically, diffusiophoresis relies on solute concentration gradients to generate  a force on colloidal particles, which is the sum of chemiphoresis and electrophoresis contributions.
Chemiphoresis takes its origin in the concentration gradient of solutes that applies a non-uniform pressure on the thin interfacial layer (\textit{e.g.} Debye layer) and electrophoresis arises from a local electric field created by the difference of diffusivities between anions and cations.
Recently, gels have been used as a steady source of solute over long time scales to get a long lasting electrophoresis effect \cite{Banerjee2016}.

In previous studies, we reported observations and analyzed the absorption of a drop \cite{Kajiya2011} and the particle deposition  \cite{Boulogne2015b,Boulogne2016b}  on swelling hydrogels.
Particles are transported along with the solvent such that micron size particles are deposited at the interface of the hydrogel to form a nearly uniform coating \cite{Boulogne2015b}.
Furthermore, we have shown that the radial profile of the number of deposited particles can be controlled by the particle concentration \cite{Boulogne2016b}.

In this paper, we combine diffusiophoresis as a particle transport mechanism and the porous properties of hydrogels as a solute source/sink.
More precisely, we investigate the effect of a difference of solute concentration between the colloidal drop and the hydrogel.
We show that a gradient of solute concentration enables the manipulation of microspheres in the drop, either to repel them from the gel surface or to generate adsorption depending on the surface charge of the particles.
Our aim is to show how the well-studied diffusiophoresis mechanism can be used for the deposition of particles on solids from dispensed drops, beyond the classic evaporation technique or the solvent absorption by the substrate.

\section{Experimental details}

\begin{figure}
    \centering
    \includegraphics[width=\linewidth]{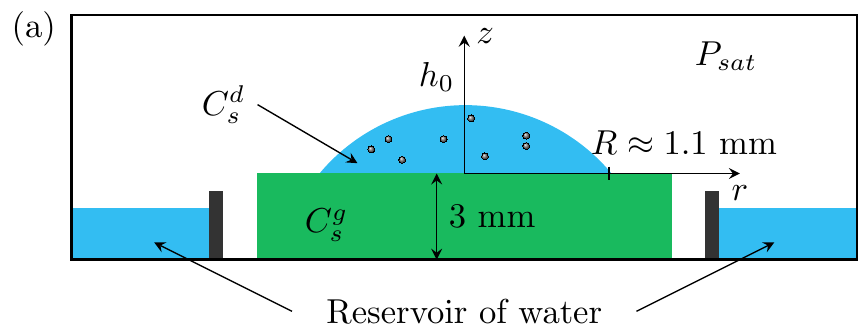}\\
    \includegraphics[width=\linewidth]{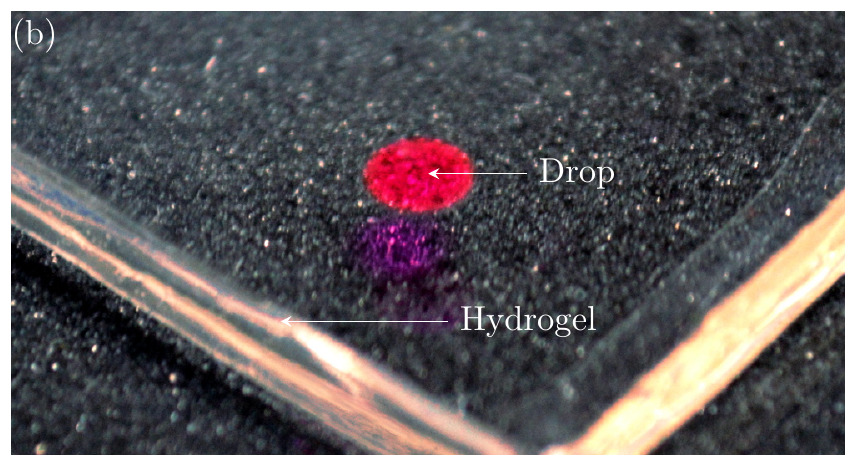}
    \caption{(a) Schematic of the experimental setup. $C_s^g$ denotes the initial solute concentration in the hydrogel and $C_s^d$ denotes the initial concentration in the drop.
    A sealed box with a reservoir of water prevents the drying of the drop and the hydrogel by saturating the atmosphere.
    (b) Picture of a drop  deposited on a slab of hydrogel.
    The drop is colored red for illustrative purpose.
    }\label{fig:setup}
\end{figure}

The substrate is a hydrogel of $3$ mm thickness fully swollen with an ionic solution of sodium chloride (NaCl) at a concentration $C_s^g$.
A drop containing fluorescent microspheres in an ionic solution of concentration $C_s^d$ of the same solute is deposited on the hydrogel.
The system is placed in a sealed box saturated in water vapor to prevent evaporation (Fig. \ref{fig:setup}).
In the present conditions, since the hydrogel is initially fully swollen, the drop is not absorbed by the gel in contrast to our previous studies \cite{Boulogne2015b,Boulogne2016b}.

\subsection{Hydrogels}

The gel is a polydimethyl acrylamide (PDMA) hydrogel \cite{Sudre2011}.
To prepare these hydrogels, we used a monomer N,N-dimethylacrylamide (DMA), a crosslinker N,N'-methylene-bis-acrylamide (MBA)
and initiators, potasium persulfate (KPS) and N,N,N',N'-tetramethylethylenediamine (TMED).
All chemicals are purchased from Sigma-Aldrich, USA.
The composition of these hydrogels is the same as some of those used in our previous study on the deposition by solvent absorption \cite{Boulogne2015b}.

In addition, $[\rm{MBA}]$ and $[\rm{mono}]$ are, respectively, molar concentrations of the crosslinker (N,N'-methylene-bis-acrylamide) and the monomer.
The composition of the gel is set by $m_{\rm DMA} / (m_{\rm DMA} + m_{w}) = 0.15$
whereas $m_{\rm DMA}$ and $m_w$ are the masses of DMA and water, respectively.
The molar concentration ratio of MBA and DMA is set to $[\rm{MBA}]/[\rm{DMA}] = 0.02$.
The quantities of initiators KPS and TMED are set to a molar ratio of 1\% of the monomer quantity.
The solution is poured in a mold made of two glass plates ($75\times50$ cm$^2$, Dow Corning) separated with a rubber spacer of $3$ mm thickness (McMaster-Carr).
Hydrogels are stored in their molds in a vapor saturated environment for two days.

After removing the mold, hydrogels are dialysed in deionized water for five days.
During this process, the water bath is changed twice and it results in a fully swollen gel.
These gels are then transferred into NaCl solution at a concentration of $0.1$ mM or $10$ mM for at least five days.
At the end of this process, we assume that the hydrogels have a solute concentration $C_s^g$ equal to the concentration of the NaCl solution.

Prior to each experiment, a gel is removed from the ionic bath and the surface is gently wiped with a weighing paper (Fisherbrand) to remove the excess solution.
The gel is placed in a hermetic glass cell containing a reservoir of water to saturate the gas phase and to prevent evaporation of the drop and the gel.
A drop of colloidal suspension is dispensed on the hydrogel with a micropipette (Research Plus, Eppendorf).

\subsection{Drop of colloidal suspension}

Colloidal suspensions consist of hydrophilic fluorescent particles  (carboxylate-functionalized polystyrene, Lifetechnologies, Orange (540/560)) of diameter $2a = 1.1$ $\mu$m and with a density 1.05 g/cm$^3$ purchased as an aqueous suspension containing 0.02 \% thimerosal.
Ionic solutions of NaCl are prepared at concentrations of $C_s^d = 0.1$ mM and $10$ mM.
The suspension is diluted with the ionic solution at a concentration $C_p^0 = 8\times 10^{6}$ particles/mL, which corresponds to a volume fraction of $6\times 10^{-6}$.
The viscosity of the solvent is $\eta_s=10^{-3}$ Pa$\cdot$s.
The zeta potential of the particles is measured as $\zeta = -70$ mV using dynamic light scattering (Malvern Zetasizer Nano).
In all experiments, the drop volume is $V_d=0.8$ $\mu$L, which when placed on the gel results in a drop radius $R\approx 1.1$ mm and a drop height $h_0 \approx 0.4$ mm.

\section{Governing equations}\label{sec:theory}

In this section, we present the governing equations describing the motion of colloidal particles by diffusiophoresis.
We define successively the geometry of the system, the diffusion equation for ions, the diffusiophoretic mobility and the equation for the particle motion.
Note that, in the present work, we neglect diffusioosmotic flow driven by the surface charge of the hydrogel due to the small zeta potential \cite{Yezek2004}.

\subsection{Drop shape}

We define the capillary length $\ell_c = \sqrt{\gamma/(\rho g)}$ where $\gamma$ is the surface tension, $\rho$ the liquid density and $g$ the gravitational acceleration.
For a drop radius $R < \ell_c$, to a good approximation the drop shape is a spherical cap.
Thus, at each time, the drop profile $h(r,t)$ represented in Fig. \ref{fig:setup}(a) is described by a portion of a sphere,

\begin{equation}\label{eq:HOMO_drop_shape_general}
    h(r,t) = \sqrt{ \left(  \frac{R^2 + h_0(t)^2}{2h_0(t)} \right)^2 - r^2 }  - \frac{(R^2 - h_0(t)^2)}{2h_0(t)},
\end{equation}
with $h_0(t) = h(0,t)$.
We assume that the drop height is much smaller than the drop radius, $h_0\ll R$, which corresponds to small contact angles.
Thus, equation (\ref{eq:HOMO_drop_shape_general}) becomes

\begin{equation}\label{eq:drop_shape}
    h(r) \simeq h_0\, \left( 1 - \frac{r^2}{R^2} \right).
\end{equation}

\subsection{Ionic diffusion}

As the drop and the gel have different initial NaCl concentrations, ions diffuse through the gel-drop interface.
In cylindrical coordinates, the concentration field $c(r, z, t)$ is a solution of the time-dependent diffusion equation

\begin{equation}\label{eq:diffusion}
    \frac{\partial c}{\partial t} = {\cal D}_s \left( \frac{\partial^2 c}{\partial r^2} + \frac{1}{r} \frac{\partial c}{\partial r} + \frac{\partial^2 c}{\partial z^2} \right),
\end{equation}
where ${\cal D}_s$ is the ambipolar diffusion coefficient for the solute defined as ${\cal D}_s =  2{\cal D}_+ {\cal D}_-/( {\cal D}_+ + {\cal D}_- )$, where ${\cal D}_+$ and ${\cal D}_-$ are the diffusivities of the cation and anion, respectively.
For NaCl, the diffusion coefficient of cations is ${\cal D}_+ = 1.33\times 10^{-9}$ m$^2$/s and for anions, ${\cal D}_- = 2.03 \times 10^{-9}$ m$^2$/s \cite{Cussler2009}.
Hence, ${\cal D}_s = 1.6\times 10^{-9}$ m$^2$/s.
The initial conditions are $c(r,z>0, t=0) = C_s^d$ and $c(r,z<0, t=0) = C_s^g$.

\subsection{Diffusiophoretic mobility}

The diffusiophoretic mobility ${\cal D}_{dp}$ of the particles relates the particle velocity $\mathbf{v}_p$ due to diffusiophoresis and the solute gradient as \cite{Anderson1984,Prieve1984}

\begin{equation}\label{eq:dp_velocity}
    \mathbf{v}_p={\cal D}_{dp}\mathbf{\nabla}\ln c,
\end{equation}
where $c$ is the solute concentration. Under the thin Debye layer approximation, the diffusiophoretic mobility is given by \cite{Prieve1984}

\begin{equation}\label{eq:dp_mobility}
    {\cal D}_{dp}=\frac{\epsilon}{\eta_s}\left(\frac{k_BT}{Ze}\right)^2 \left[\frac{Ze\zeta \beta}{k_BT} + 4\ln \cosh \left(\frac{Ze\zeta}{4k_BT}\right) \right],
\end{equation}
where $\epsilon$ is the permittivity of the medium, $k_B$ is the Boltzmann constant, $T$ is the temperature, $Z$ is the valence of the solute, $\zeta$ is the zeta potential, and $\beta=({\cal D}_+ - {\cal D}_-)/({\cal D}_+ + {\cal D}_-) = -0.21$ for NaCl.

The diffusiophoretic mobility consists of electrophoretic and chemiphoretic terms, which are, respectively, the first and the second terms in the bracket on the right-hand side of equation (\ref{eq:dp_mobility}).
Electrophoresis is driven by the diffusion potential that is set by the difference in the diffusivities between cation and anion whereas chemiphoresis originates from the osmotic pressure difference within the Debye layer due to a solute gradient.

The diffusiophoretic mobility is determined from the zeta potential of the colloidal particle, which is measured as $-70$ mV.
Using equation (\ref{eq:dp_mobility}), the calculated diffusiophoretic mobility is ${\cal D}_{dp} = 7.44 \times 10^{-10}$ m$^2$/s, where for the conditions of our experiments the electrophoretic contribution is $2.95 \times 10^{-10}$ m$^2$/s and the chemiphoretic contribution is $4.49  \times 10^{-10}$ m$^2$/s.

The particle motion is generated by  the diffusiophoretic velocity given by equation (\ref{eq:dp_velocity}) and the Brownian motion associated with the particle diffusivity ${\cal D}_p$.
The diffusion coefficient of particles is obtained from the Stokes-Einstein relation ${\cal D}_p = \frac{k_B T}{6\pi\eta_s a} = 4.0 \times 10^{-13}$ m$^2$/s.
Therefore, the particle concentration $C_p(r, z, t)$ in the drop is given by
\begin{equation}\label{eq:particle_density}
    \frac{\partial C_p}{\partial t} + \mathbf{\nabla}\cdot (\mathbf{v}_p C_p) = {\cal D}_p \nabla^2 C_p,
\end{equation}
where the initial condition is $C_p(r, z, t=0) = C_p^0$.

\section{Experimental observations}

To describe the different solute concentration gradients, we define the solute concentration ratio as
\begin{equation}\label{eq:def_gamma}
    \Gamma = C_s^g/C_s^d,
\end{equation}
where $C_s^g$ and $C_s^d$ are, respectively, the initial concentrations in the gel and in the drop.
For different solute concentration ratios $\Gamma$, we analyze the particle motion in the drop.
To achieve this analysis, we combined two techniques.
First, confocal microscopy is used to visualize the particle positions in the cross-section of the drop.
Then, we observed the particles deposited on the gel interface with fluorescence microscopy.

\subsection{Particle distribution in the drop cross-section}

\begin{figure*}
    \includegraphics[scale=1]{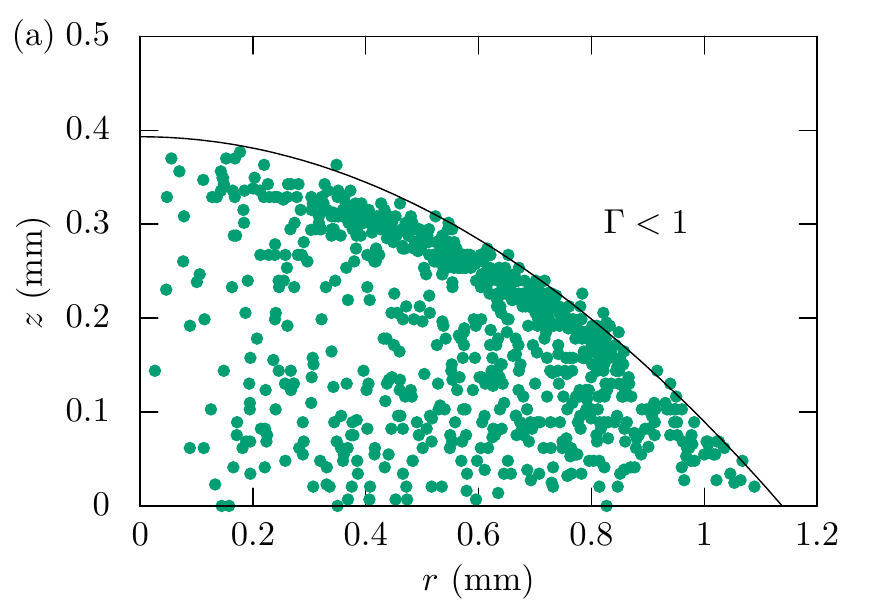}
    \includegraphics[scale=1]{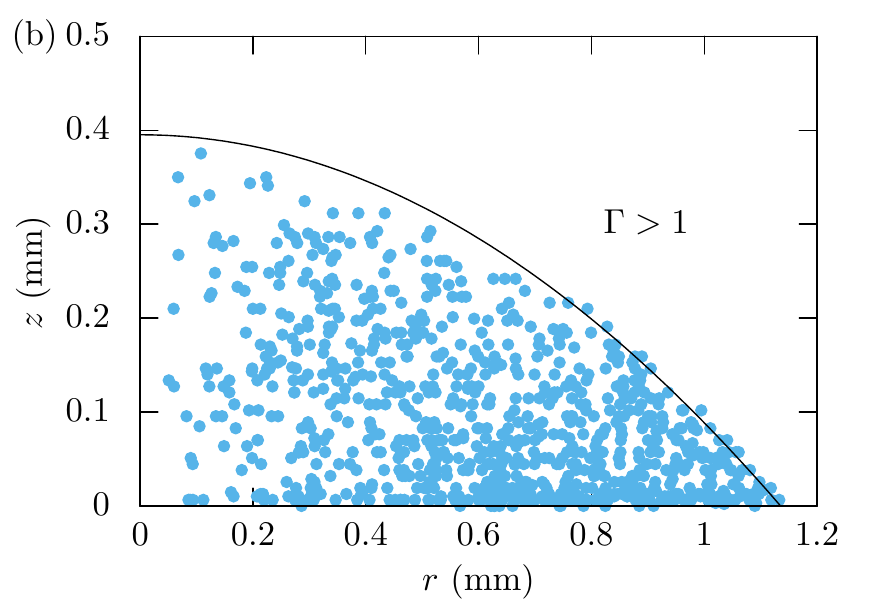}\\
    \caption{Position of particles in the drop about 4 minutes after the dispensing.
        The solid line is the drop interface given by equation (\ref{eq:drop_shape}).
        (a) $C_s^g=0.1$ mM, $C_s^d=10$ mM ($\Gamma = 0.01 < 1$)
        (b) $C_s^g=10$ mM, $C_s^d=0.1$ mM ($\Gamma = 100 > 1$).
    }\label{fig:confocal_drop}
\end{figure*}

To measure the particle position, we use a confocal microscope (Leica TCS SP5) with a 10$\times$ dry objective (PL Fluotar).
At this magnification, we image a quarter of a 0.8 $\mu\ell$ drop.
Images in the horizontal $(x,y)$-frame are taken with a $z$-step of 10 $\mu$m with a pinhole size set to $70$ $\mu$m.
The image resolution is $1024\times1024$ pixels and we used a scanning rate of $200$ Hz.
To visualize the full cross-section of the drop, we choose a $z$-range larger than the height of the droplet, which is typically $h_0\approx 0.4$ mm.
Images are recorded from the fluorescence and the transmission signals.
We start the acquisition about 2 minutes after the drop deposition and the duration of the acquisition is about 3 minutes, such that early time dynamics is not accessible.
The initial position of the scanner is located near the surface of the gel and the scanner moves upward during the acquisition.

\begin{figure}
	\centering
    \includegraphics[width=0.90\linewidth]{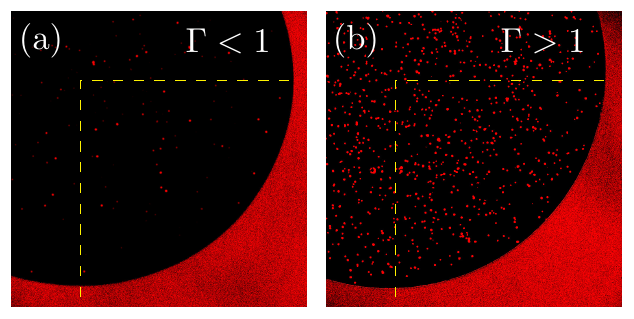}\\
    \includegraphics[scale=1]{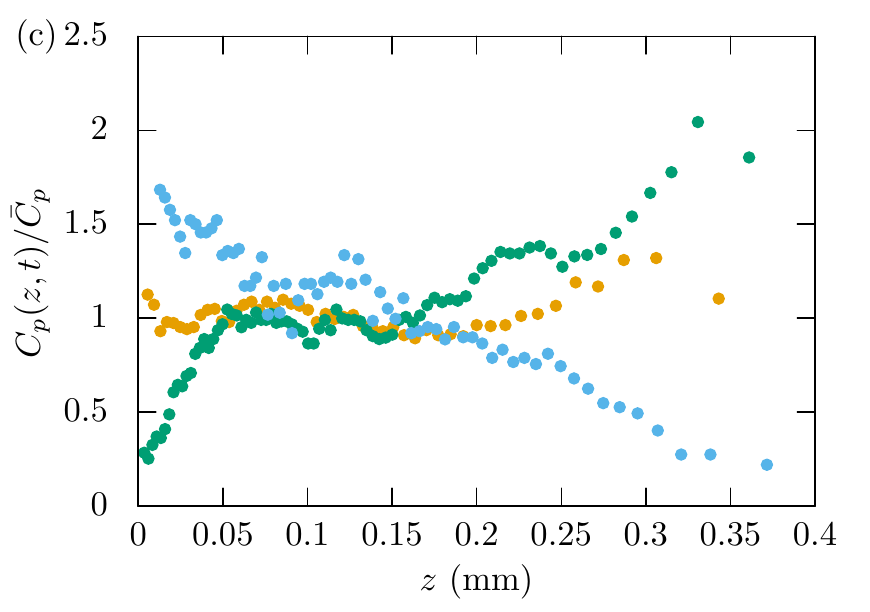}
    \caption{(a-b) Confocal images captured at the level of the hydrogel interface for (a) $\Gamma= 0.01$ and (b) $\Gamma = 100 $.
    We observe the interface reflection outside the drop.
    The dashed lines show the quarter of the drop used in the analysis.
        (c) Particle concentration profile for three solute concentration ratios $\Gamma$ captured typically between 2 to 5 minutes after deposition.
    Green points correspond to the data presented in Fig. \ref{fig:confocal_drop}(a), $\Gamma =0.01$, and blue points to Fig. \ref{fig:confocal_drop}(b), $\Gamma = 100$.
    Orange points are for $C_s^g=C_s^d = 0.1$ mM, \textit{i.e.} $\Gamma = 1$.
    }\label{fig:confocal_profile}
\end{figure}

From a $z$-stack, we start by detecting the gel interface and the contact line of the drop.
From a region where the gel-vapor interface is visible on  pictures, we measure the average fluorescence intensity as a function of the $z$ position.
With the interface reflectivity, the peak of intensity corresponds to the position of the gel interface.
On the corresponding image in transmission mode, the position of the contact line is detected by a circular Hough transform \cite{Vanderwalt2014}.
As the gel and the drop are now located, we focus on the detection of the particles.

A particle can be seen on several images of the $z$-stack with a displacement in the $(x,y)$ frame due to Brownian motion occurring during the acquisition.
To determine an average position, we use the Python library Trackpy \cite{Allan2014} along with the scientific libraries numpy, scipy, pandas \cite{Oliphant2007,Millman2011,Walt2011,McKinney2010}.
On the fluorescence images, the particle $(x,y)$ positions are tracked in the $z$-stack.
From the resulting particle trajectories, we calculate the average $x$, $y$ and $z$ positions for each particle.
The total number of particles in the entire image corresponds to $C_p V_d /4$ to a precision of $15$\%.

\begin{figure*}
    \includegraphics[scale=1]{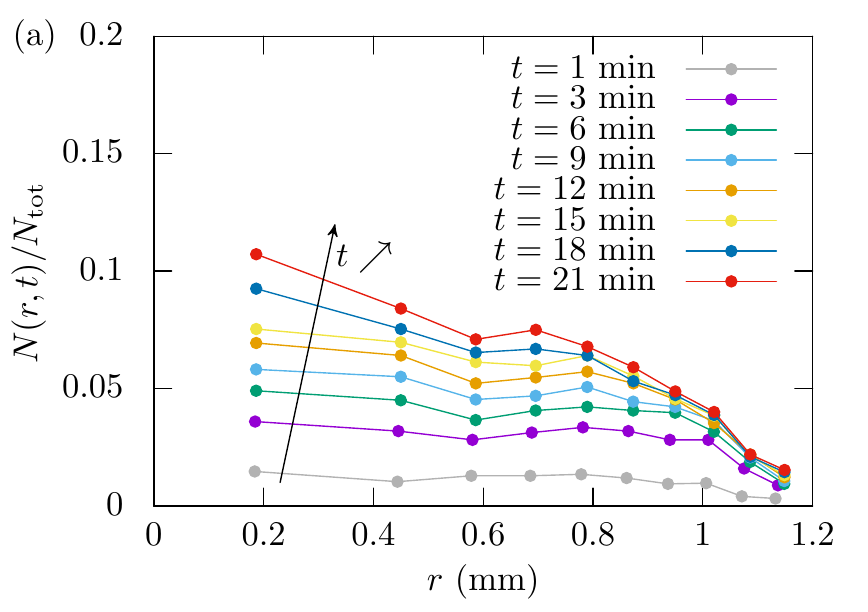}
    \includegraphics[scale=1]{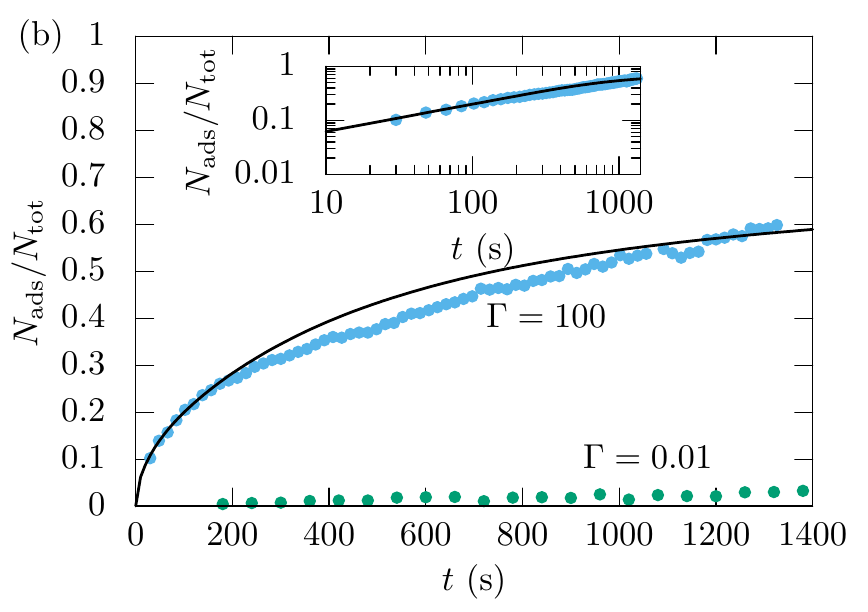}
    \caption{(a) Radial profiles of the particle deposited on the hydrogel ($\Gamma = 100$) for times spanning from 1 to 21 min obtained from fluorescence microscopy.
        The profiles are non-dimensionalized by the total number particles $N_{{\rm tot}}$ initially present in the drop.
        (b) Number of adsorbed particles $N_{{\rm ads}}$, non-dimensionalized by the total number of particles $N_{{\rm tot}}$.
        Green points are for $C_s^g=0.1$ mM and $C_s^d=10$ mM ($\Gamma=0.01$)
        and blue points are for $C_s^g=10$ mM and $C_s^d=0.1$ mM ($\Gamma = 100$).
        The solid black line is the numerical solution with a height $h_0^{(n)}$ adjusted at 1.2 mm as a fitted parameter.
    }\label{fig:fluo2_N_time}
\end{figure*}

In Fig. \ref{fig:confocal_drop}, we show the particle positions for two solute concentration ratios.
Since the particles are detected in one quarter of the drop, the liquid volume in a slice ${\rm d}r$ is $h(r)\frac{\pi}{2} r {\rm d} r$.
Therefore, at a given position $z$ in the drop, fewer particles are visible as $r \rightarrow 0$ in this 2D representation.
For larger solute concentrations in the drop, \textit{i.e.} $\Gamma < 1$, and for this choice of salt (NaCl), we observe that the particles are repelled from the gel interface and they accumulate near the liquid-air interface (Fig. \ref{fig:confocal_drop}(b)).
In contrast, for $\Gamma > 1$, the particles migrate toward the hydrogel (Fig. \ref{fig:confocal_drop}(b)).

The number of particles ${\rm d}n(z, t)$ contained in a horizontal slice of liquid of thickness ${\rm d}z$ is related to the concentration $C_p(z, t)$ by
\begin{equation}
    {\rm d}n(z, t) = C_p(z, t)\,\pi\,r(z)^2 {\rm d} z.
\end{equation}
In practice, we count the number of particles located in horizontal cross-sections of constant volume to ensure that a sufficient number of particles are present in each elementary volume.
In Fig. \ref{fig:confocal_profile}(c), we plot the relative particle concentration profile $C_p(z, t)/\bar{C}_p$, where $\bar{C}_p$ represents the average particle concentration.
For $\Gamma=1$ (orange points), the solute concentration is uniform everywhere, thus no diffusiophoretic effect is expected and the experimental concentration profile is nearly flat.

For $\Gamma > 1$, \textit{i.e.} a higher solute concentration in the gel than the drop, the experimental results show that the concentration near the surface of the hydrogel is about twice the average concentration.
For the opposite gradient $\Gamma < 1$, particles are significantly repelled from the gel surface with a local concentration of about 30 \% the initial concentration \cite{Musa2016}.
Note that the direction of the particle motion is independent of the sign of the zeta potential since chemiphoresis dominates over electrophoresis for NaCl solution within a wide range of zeta potentials \cite{Shin2017b}. 
This also implies that the particle diffusiophoretic motion can be reversed for positively-charged particles when an electrophoresis-dominant solute is used \cite{Shin2017}.
In the next section, we focus our attention on the particle deposition rate.

\subsection{Dynamics of the particle deposition on  the hydrogel interface}

To identify the particles adsorbed on the hydrogel, we performed experiments with an inverted fluorescent microscope (Leica DMI4000 B) with a $5\times$ objective.
The microscope is equipped with a translation stage (Prior) and a Hamamatsu camera (Digital camera ORCA-Flash4.0 C11440) $2048\times 2048$ pixels.
The acquisition is automated with the software Micromanager \cite{Edelstein2010}.

The microscope is focused on the top surface of the hydrogel.
An acquisition consists in a sequence of 5 images captured in the fluorescence mode followed by an image in bright field.
The drop position and its radius are measured with a circular Hough transform as before.
With the library Trackpy, we capture the particle trajectories over a set of 5 images.
Among these trajectories, we define a criterion to determine if a particle is adsorbed on the substrate, \textit{i.e.} not mobile.
For each trajectory consisting of a set of positions $(r_i, \theta_i)$ in the horizontal plane,  we define $\bar{r}$, the average radial position and $\textrm{std}()$, the standard deviation.
Then, we calculate the radial displacement $\delta r = \textrm{std}(r_i)$ and the azimuthal displacement $\bar{r} \delta \theta= \bar{r}\, \textrm{std}(\theta_i)$.
We consider that a particle is not mobile if its trajectory satisfies $\delta r < 4$ $\mu$m and $\bar{r} \delta \theta < 4$ $\mu$m.
This procedure is repeated over the sequences.
The diffusion timescale for the particles over $\delta r$ is $(\delta r)^2 / {\cal D}_p \approx 1$ s.
The time step between each picture is $3$ s, such that the duration of a sequence is 15 s, which is much larger than the diffusion timescale to ensure a correct detection of adsorbed particles.

We represent the time evolution of the radial particle distribution in Fig. \ref{fig:fluo2_N_time}(a).
At early times ($t\lesssim 100$~s), the radial profile is nearly flat  and as time passes, the density increases more at the center than near the contact line.
Indeed, we expect that the solute gradient is mainly vertical, such that the main velocity component of the particles is also vertical as represented by equation (\ref{eq:dp_velocity}).
Therefore, fewer particles tend to accumulate near the drop edge and there is a decreasing concentration profile along the radius.
Interestingly, we noticed that the mobile particles in the drop have an inward radial motion on the first minutes after the drop deposition.

Next, we plot the total number of absorbed particles as a function of time for $\Gamma=0.01$ and $\Gamma=100$, as shown in Fig. \ref{fig:fluo2_N_time}(b).
For $\Gamma = 0.01$, we observe a weak particle absorption of only a few percent at the end of the experiment.
Therefore, most of the particles are repelled from the gel surface and the few adsorbed particles might be due to the irreversible adsorption when particles move near the surface.
The good adhesion of these particles on PDMA hydrogels has been noticed in our previous studies \cite{Boulogne2015b,Boulogne2016b} where we reported that a contact line cannot detach adsorbed particles.
For $\Gamma = 100$, the number of adsorbed particles increases non-linearly in time, with a decreasing absorption rate.
In the next section, we solve numerically a simplified model based on the governing equations presented in section \ref{sec:theory} for the case $\Gamma>1$ to rationalize our data.

\section{Model}

\begin{figure}
    \includegraphics[width=\linewidth]{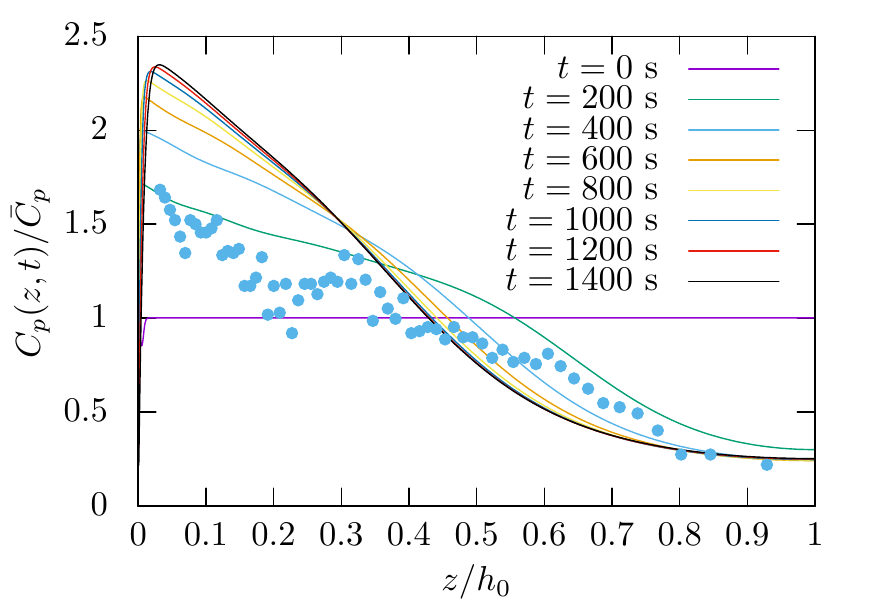}\\
    \caption{Normalized particle concentration as a function of the rescaled drop height $z/h_0$.
    Solid lines are the concentration profiles obtained numerically for an effective drop height $h_0^{(n)}=1.2$ mm at different times.
    Points are the experimental data presented in Fig.~\ref{fig:confocal_profile}(c) for $\Gamma= 100$, rescaled by the drop height $h_0=0.4$ mm.
    The acquisition of the experimental data started 2 minutes after the drop deposition and the duration is about 3 minutes.
    }\label{fig:model}
\end{figure}

The derivation of an analytical solution is particularly difficult because of the finite size effect due to the limited drop height.
We make the choice to simplify the geometry of the system to retain the essential physical behavior.
Thus, we consider a 1D model where the drop is approximated as a thin cylinder of height $h_0^{(n)}$.
The governing equations (\ref{eq:diffusion}), (\ref{eq:dp_velocity}) and  (\ref{eq:particle_density})  can be rewritten for $c(z,t)$ as

\begin{subequations}\label{eq:master_equation_model}
\begin{align}
    \frac{\partial c}{\partial t} &= {\cal D}_s \frac{\partial^2 c}{\partial z^2},\label{eq:diffusion_1D}\\
    v_p&= {\cal D}_{dp}\frac{\partial{\ln c}}{\partial z}, \label{eq:dp_velocity_1D}\\
    \frac{\partial C_p}{\partial t} &= - \frac{\partial (C_p v_p)}{\partial z} + {\cal D}_p \frac{\partial^2 C_p}{\partial z^2},\label{eq:particle_density_1D}
    \end{align}
\end{subequations}
where $C_p(z,t)$ is the particle concentration in the drop.
We define diffusion timescales $\tau = h_0^2 / {\cal D}$  across the drop height $h_0$ for the three diffusion coefficients in the problem, namely ${\cal D}_s$, ${\cal D}_p$ and ${\cal D}_{dp}$.
For a drop height $h_0\approx 0.4$ mm, we have $\tau_s \approx 100$ s,  $\tau_p \approx 4\times 10^5$ s and $\tau_{dp} \approx 200$~s.
The shorter two times indicate that at the timescale of the experiments, the finite size of the droplet matters.
Hence, we next consider boundary conditions accounting for the finite geometry.

We consider the case $\Gamma > 1$ corresponding to the experimental observation of migration of the particles toward the gel interface.
Because the volume of the drop is small compared to the volume of the gel, we assume that the migration of solute molecules from the gel toward the drop does not appreciably change the concentration of solute inside the gel.
Furthermore, we shall assume that colloidal particles are irreversibly adsorbed when they reach the drop/gel interface, \textit{i.e.} $n(z=0, t)=0$, such that the interface behaves as a \textit{perfect sink} \cite{Bacchin1995}.
In addition, neither solutes nor colloidal particles can cross the drop/air interface and thus, for a 1D drop with height $h_0^{(n)}$, the boundary conditions read:

\begin{subeqnarray}
    c(z=0,t) &=& C_s^g,\\
    C_p(z=0,t) &=& 0,\\
    \frac{\partial c}{\partial z}\bigg\rvert_{z=h_0^{(n)},t} &=& 0,\\
   {\cal D}_p \frac{\partial C_p}{\partial z}\bigg\rvert_{z=h_0^{(n)},t} &=& C_p\left(h_0^{(n)},t\right)v_p\left(h_0^{(n)},t\right),
   \label{eq:BC}
\end{subeqnarray}

\noindent together with the initial conditions:

\begin{subeqnarray}
c(z, t=0) &=& C_s^d,\\
C_p(z, t=0) &=& C_p^0.
\label{eq:IC}
\end{subeqnarray}

We then solved numerically equations \ref{eq:master_equation_model}(a-c), subjected to boundary and initial conditions (\ref{eq:BC}) and (\ref{eq:IC}).
In Fig.~\ref{fig:fluo2_N_time}(b), since the model assumes a simpler geometry, we fit our data using the thickness $h_0$ of the cylindrical drop as a free parameter.
The best fit is obtained for $h_0^{(n)}=1.2$ mm, which is of the same order of magnitude as the experimental value $h_0 \approx 0.4$~mm.
Given the highly simplified geometry used in this model, this estimate is reasonable.
Similar agreement for the time-dependent dynamics can be obtained analytically with further simplifications (see Appendix).

In Fig.~\ref{fig:model}, we show the numerical prediction of the particle concentration profiles at different times ranging from 0 to 1400~s.
The curves indicate a depletion effect near the interface as a result of the \textit{perfect sink} boundary condition that removes the particles from the liquid phase as they irreversibly adsorb.
This depleted zone has a maximum characteristic thickness of $0.03\,h_0^{(n)}$.
Then, the concentration monotonically decreases along the drop height as the particles migrate toward $z=0$ in time.
The data points represented in Fig.~\ref{fig:model} are extracted from Fig.~\ref{fig:confocal_profile}(c) and the acquisition has been performed between $t=120$~s and $t=300$~s.
The numerical model predicts rather accurately the vertical particle distribution, as can be seen in Fig. \ref{fig:model} (green curve).
The size of the depleted zone is clearly outside of the experimental resolution such that it is not surprising that it has not been observed.

Further work would be necessary to predict the radial distribution of deposited particles as shown in Fig.~\ref{fig:fluo2_N_time}(a).
This would require to solve the set of equations (\ref{eq:diffusion}), (\ref{eq:dp_velocity}) and (\ref{eq:particle_density}) with the geometry of the drop given by equation (\ref{eq:drop_shape}).
In particular, such description would capture the significant finite-size effect occurring in the wedge formed by the contact line.
Also, the solution for the 3D salt concentration field would probably suggest an outward diffusive flux in the drop responsible for the radial inward motion of particles that we observed after the drop deposition.

\section{Conclusion}

We studied particle transport induced by diffusiophoresis in a water drop deposited on a hydrogel.
We show that a difference of solute concentration between the hydrogel and the drop triggers either a particle motion towards the gel interface or, on the contrary, toward the liquid-air interface of the drop, depending on the direction of the solute gradient.
This behavior can be understood qualitatively by inspection of the equations describing the diffusion of solute and the diffusiophoretic mobility.
In the particular situation of particle deposition, which is our interest, we observed that the density profile decreases along the radius.
This trend is more pronounced as the deposition by diffusiopheresis proceeds.
In addition, we derived a simplified 1D numerical model that successfully rationalizes our findings for the time evolution of the total number of adsorbed particles.

We believe that the results reported in this paper are important for at least two reasons.
First, when coating the surface of a hydrogel with colloidal particles, not only solvent evaporation or absorption can be used;
diffusiophoretic forces allow manipulation of the colloidal particles on timescales that are comparable to the deposition made by absorption that we reported previously \cite{Boulogne2015b,Boulogne2016b}.
Furthermore, the direction of the solute gradient can be chosen to manipulate the particles either toward the solid interface or toward the liquid-air interface, depending on the surface charge of the particles.
Our experiments use hydrogels but these conclusions must remain relevant for various porous substrates.
The second message is to bring attention to solute concentration gradients in situations where colloidal particles and hydrogels are manipulated simultaneously.
Indeed, solute concentration gradient may appear due to the presence of ions in the colloidal suspension or in the hydrogel.
Whereas we have not observed a particle motion that can be attributed to diffusiophoresis in our previous experiments \cite{Boulogne2015b,Boulogne2016b}, the presence of ions must be carefully considered as they can generate significant effects as we have shown in this paper.

\section{Acknowledgments}
We kindly thank Fran\c{c}ois Ingremeau for useful discussions.
F.B. acknowledges that the research leading to these results received funding from the People Programme (Marie Curie Actions) of the European Union's Seventh Framework Programme (FP7/2007-2013) under REA grant agreement 623541.
S.S. and H.A.S. thank Unilever Research and the Princeton Environmental Institute for partial support of this research.

\bibliography{article_diffusiophoresisgel}
\bibliographystyle{unsrt}
\newpage\clearpage

\section{Appendix: a simple analytical model}

We derive in this section a simple model for the case $\Gamma > 1$ that can be used to provide some insights on the dynamics of colloids deposition at the surface of hydrogels. For the sake of simplicity, let us neglect at first the finite size of the drop. Under this crude approximation the solute distribution is self-similar and can be written as \cite{Crank1975}:

\begin{equation}
\label{eq:solution_salt}
 \frac{c(z,t) - C_s^d}{C_s^g - C_s^d} = {\rm erfc}\left(\frac{z}{2\sqrt{{\cal D}_s t}} \right).
\end{equation}
As the finite size effects are neglected, this equation is valid only for $t \ll h_0^2 / {\cal D}_s$.

Substituting equation (\ref{eq:solution_salt}) in equation (\ref{eq:dp_velocity_1D}), we obtain the following expression for the diffusiophoretic velocity:

\begin{equation}
\label{eq:vp_sol1}
    v_p(z, t) = - \frac{{\cal D}_{dp}}{\sqrt{\pi {\cal D}_s t}} \frac{e^{-\chi^2}}{{\rm erfc}(\chi) }, \,\,\, \mbox{where} \,\,\, \chi = \frac{z}{2 \sqrt{{\cal D}_s t}}.
\end{equation}

Here, $v_p$ is negative as the particles are moving toward the gel interface for the initial condition we consider.
At the gel interface, we have:

\begin{equation}
    v_p(z=0, t) = - \frac{{\cal D}_{dp}}{\sqrt{\pi {\cal D}_s t}}.
\end{equation}

At a vertical position corresponding to the height $h_0$ of the drop, we can estimate the factor $e^{-\chi^2} / {\rm erfc}(\chi)$ in equation (\ref{eq:vp_sol1}) for different times that are accessible experimentally:

\begin{equation}
    v_p(z=h_0, t) \approx - \frac{{\cal D}_{dp}}{\sqrt{\pi {\cal D}_s t}} \left\{
        \begin{array}{ll}
            \times 3.0 & \mbox{for t=10 s,}  \\
            \times 1.6 & \mbox{for t=100 s,}\\
            \times 1.2 & \mbox{for t=1000 s.}\\
        \end{array}
    \right.
\end{equation}
We notice that for the range of timescales and vertical positions that are of interest, the factor $e^{-\chi^2}/{\rm erfc}(\chi)$ is a weak function of $z$ and $t$.
Therefore, from equation (\ref{eq:vp_sol1}), we make the following approximation for the diffusiophoretic particle velocity:

\begin{equation}\label{eq:vp_sol2}
    v_p(z=h_0, t) \approx - \frac{{\cal D}_{dp}}{\sqrt{\pi {\cal D}_s t}}.
\end{equation}
It is important to stress that this approximation is not good for $t\rightarrow 0$.
Now, we consider equation (\ref{eq:particle_density_1D}) for the particle motion which, combined with (\ref{eq:vp_sol2}), leads to:
 \begin{equation}\label{eq:edp}
     \frac{\partial C_p}{\partial t} -  \frac{{\cal D}_{dp}}{\sqrt{\pi {\cal D}_s t}} \frac{\partial C_p}{\partial z} = {\cal D}_p  \frac{\partial^2 C_p}{\partial z^2},
 \end{equation}

We define $\xi = z - \frac{2 {\cal D}_{dp} }{ \pi {\cal D}_s} \sqrt{t}$ and rewrite \eqref{eq:edp} as:

\begin{equation}
\label{eq:edp2}
     \frac{\partial C_p}{\partial \xi} +  \frac{{\cal D}_{p} \sqrt{\pi {\cal D}_{s} t}}{{\cal D}_{dp}}  \frac{\partial^2 C_p}{\partial \xi^2} = 0.
\end{equation}

By integrating this equation twice, we obtain:

\begin{equation}
\label{eq:solution1_cond1}
C_p(z,t) = A \exp\left( - \frac{{\cal D}_{dp}}{{\cal D}_{p}} \frac{z}{\sqrt{\pi {\cal D}_{s} t} } \right) + B,
\end{equation}

\noindent where $A$ and $B$ are two constants. We consider the boundary conditions $C_p(z\rightarrow\infty,t) = C_p^\infty$ and $C_p(z=0,t)=0$ because the particles at the interface are not mobile and the interface behaves as a perfect sink \cite{Bacchin1995}.
Therefore, equation \eqref{eq:solution1_cond1} reduces to:

\begin{equation}
\label{eq:solution2_cond1}
C_p(z,t) = C_p^\infty \left[ 1 - \exp\left( - \frac{{\cal D}_{dp}}{{\cal D}_{p}} \frac{z}{\sqrt{\pi {\cal D}_{s} t} } \right) \right].
\end{equation}

The number of particles $N_{{\rm abs}}$ adsorbed at the interface ($z=0$) is given by:

\begin{equation}\label{eq:adv_diff}
\frac{{\rm d} N_{{\rm abs}}}{ {\rm d} t} = - C_p(z=0,t) v_p + {\cal D}_p \frac{\partial C_p}{\partial z},
\end{equation}

\noindent where, by definition of the boundary condition, $C_p(z=0,t) = 0$. In equation \eqref{eq:adv_diff}, we substitute the equation \eqref{eq:solution2_cond1}.
To take into account the finite effect of the drop, we consider that the concentration $n_\infty$ varies with the number of adsorbed particles, \textit{i.e.} $n_\infty \approx (N_{{\rm tot}} - N_{{\rm abs}}) / h_0$.
Therefore, equation \eqref{eq:adv_diff} becomes:

\begin{equation}
\frac{{\rm d} N_{{\rm abs}}}{{\rm d} t} =  \left(N_{{\rm tot}} - N_{{\rm abs}}\right) \frac{ {\cal D}_{dp} }{ h_0 \sqrt{\pi {\cal D}_s} t }.
\end{equation}

We integrate this equation with the initial condition $N_{{\rm abs}}(t=0) =0$ to obtain:

\begin{equation}\label{eq:N_abs_analytic}
\frac{N_{{\rm abs}}(t) }{ N_{{\rm tot}}} = \left[1 -  \exp\left( - \frac{ {\cal D}_{dp} \sqrt{t}}{h_0 \sqrt{\pi {\cal D}_s} }   \right) \right].
\end{equation}

As stated for Eq.~\ref{eq:solution_salt}, this solution is valid only for $t \ll h_0^2 / {\cal D}_s$.
In our experimental conditions, this assumption is equivalent to $t\ll h_0^2 {\cal D}_s / {\cal D}_{dp}^2$.
Thus, developing Eq.~\ref{eq:N_abs_analytic} at the first order, we obtain
\begin{equation}\label{eq:N_abs_analytic_simplified}
	\frac{N_{{\rm abs}}(t) }{ N_{{\rm tot}}} \simeq \frac{{\cal D}_{dp}}{h_0 \sqrt{\pi {\cal D}_s} }  \sqrt{t}.
\end{equation}
Despite the strong approximation on the particle velocity made at short timescales with Eq.~\ref{eq:vp_sol2}, Eq.~\ref{eq:N_abs_analytic_simplified} correctly describes our observations.
Indeed, the model predicts a prefactor $\frac{{\cal D}_{dp}}{h_0 \sqrt{\pi {\cal D}_s} } \simeq 0.025 $ s$^{-1/2}$ and a fit of the experimental data provides a value of $0.017$ s$^{-1/2}$.
However, because of its simplicity, this model cannot, in contrast to the numerical simulation presented in the main text, accurately describes the vertical particle distribution.

\end{document}